\documentclass[twocolumn, iop, times, tighten]{aastex631}

\usepackage{graphicx}
\usepackage{farhan-defs}

\usepackage[super]{nth}

\received{2024 October 16}
\revised{2025 March 11}
\accepted{2025 March 21}

\shorttitle{The Simulated and Observed Cosmic Web}
\shortauthors{Luber, Hasan, et al.}

\begin{document}

\title{CHILES IX: Observational and Simulated HI Content and Star Formation of Blue Galaxies in Different Cosmic Web Environments}

\correspondingauthor{Nick Luber}
\email{nicholas.m.luber@gmail.com}


\author{Nicholas M. Luber}
\affiliation{Department of Astronomy, Columbia University, 550 West 120th Street, New York, NY 10027, USA}

\author[0000-0002-0072-0281]{Farhanul Hasan}
\affiliation{Department of Astronomy, New Mexico State University, Las Cruces, NM 88003, USA}
\affiliation{Space Telescope Science Institute, 3700 San Martin Drive, Baltimore, MD 21218, USA}

\author[0000-0002-7679-9344]{J. H. van Gorkom}
\affiliation{Department of Astronomy, Columbia University, 550 West 120th Street, New York, NY 10027, USA}

\author[0000-0001-7996-7860]{D.J. Pisano}
\affiliation{Department of Astronomy, University of Cape Town, Private Bag X3, Rondebosch 7701, South Africa}

\author[0000-0002-1979-2197]{Joseph N. Burchett}
\affiliation{Department of Astronomy, New Mexico State University, Las Cruces, NM 88003, USA}


\author[0000-0002-9627-7519]{Julia Blue Bird}
\affiliation{National Radio Astronomy Observatory, P.O. Box O, Socorro, NM 87801, USA}

\author[0000-0003-1436-7658]{Hansung B. Gim}
\affiliation{Department of Physics, Montana State University, P.O. Box 173840, Bozeman, MT 59717, USA}

\author[0000-0001-9662-9089]{Kelley M. Hess}
\affiliation{Department of Space, Earth and Environment, Chalmers University of Technology, Onsala Space Observatory, 43992 Onsala, Sweden}
\affiliation{ASTRON, the Netherlands Institute for Radio Astronomy, Postbus 2, 7990 AA, Dwingeloo, The Netherlands}

\author{Lucas R. Hunt}
\affiliation{National Radio Astronomy Observatory, P.O. Box O, Socorro, NM 87801, USA}

\author[0000-0003-3385-6799]{David C. Koo}
\affiliation{UCO/Lick Observatory, Department of Astronomy and Astrophysics,
University of California, Santa Cruz, Santa Cruz, CA 95064, USA}

\author[0000-0001-6615-5492]{Sushma Kurapati}
\affiliation{Department of Astronomy, University of Cape Town, Private Bag X3, Rondebosch 7701, South Africa}
\affiliation{Netherlands Institute for Radio Astronomy (ASTRON), Oude Hoogeveensedijk 4, NL-7991 PD Dwingeloo, the Netherlands}

\author{Danielle Lucero}
\affiliation{Department of Physics, Virginia Tech, 850 West Campus Drive, Blacksburg, VA 24061, USA}

\author[0000-0001-8057-5880]{Nir Mandelker}
\affiliation{Centre for Astrophysics and Planetary Science, Racah Institute of Physics, The Hebrew University, Jerusalem 91904, Israel}

\author{Martin Meyer}
\affiliation{International Centre for Radio Astronomy Research (ICRAR), University of Western Australia, 35 Stirling Hwy, Crawley, WA 6009, Australia}

\author{Emmanuel Momjian}
\affiliation{National Radio Astronomy Observatory, P.O. Box O, Socorro, NM 87801, USA}

\author[0000-0002-6766-5942]{Daisuke Nagai}
\affiliation{Department of Physics, Yale University, New Haven, CT 06520, USA}

\author[0000-0001-5091-5098]{Joel R. Primack}
\affiliation{Department of Physics, University of California, Santa Cruz, CA 95064, USA}

\author{Min S. Yun}
\affiliation{Department of Astronomy, University of Massachusetts, Amherst, MA 01003, USA}

\begin{abstract}

We examine the redshift evolution of the relationship between the neutral atomic hydrogen ({\HI}) content and star-formation properties of blue galaxies, along with their location in the cosmic web. Using the COSMOS {\HI} Large Extragalactic Survey (CHILES) and the IllustrisTNG (TNG100) cosmological simulation, and the {\disperse} algorithm, we identify the filamentary structure in both observations and simulations, measure the distance of galaxies to the nearest filament spine {\dfil}, and calculate the mean {\HI} gas fraction and the relative specific star formation rate (sSFR) of blue galaxies in three different cosmic web environments -- $0<{\dfil}/\mathrm{Mpc}<2$ (filament cores), $2<{\dfil}/\mathrm{Mpc}<4$ (filament outskirts), and $4<{\dfil}/\mathrm{Mpc}<20$ (voids). We find that, although there are some similarities between CHILES and TNG, there exist significant discrepancies in the dependence of {\HI} and star formation on the cosmic web and on redshift. TNG overpredicts the observed {\HI} fraction and relative sSFR at $z=0-0.5$, with the tension being strongest in the voids. CHILES observes a decline in the {\HI} fraction from filament cores to voids, exactly the opposite of the trend predicted by TNG. CHILES observes an increase in {\HI} fraction at $z=0.5\rightarrow0$ in the voids, while TNG predicts an increase in this time in all environments. Further dividing the sample into stellar mass bins, we find that the {\HI} in ${\logms}>10$ galaxies is better reproduced by TNG than {\HI} in ${\logms}=9-10$ galaxies. 

\end{abstract}


\section{Introduction}\label{sec:intro}

\quad We have long observed a clear distinction between star-forming (blue) and quiescent (red) galaxies. This distinction is clearly visible through the distribution of galaxies, which is almost perfectly red-blue bimodal \citep{baldry04,faber07}, except for a small group of ``green valley" galaxies \citep{brammer09,schawinski14}. These different types of galaxies have specific physical properties that correspond to their colors, such as the color-morphology \citep{chester64,faber73,roberts94}, color-stellar mass \citep{baldry06}, and {\HI} mass-NUV-r color relationships \citep{catinella10}. The differences between these populations also extend to their environment, such as the morphology-density relationship \citep{dressler80} and relationships with the local environment as characterized by, e.g., galaxy overdensity \citep{baldry06}.

\quad Understanding the physical mechanism behind the transition of galaxies from star-forming to quiescent and the loss of their gaseous reservoirs is one of the most fundamental issues in forming a fully predictive theory of galaxy evolution. In the cluster environment, various mechanisms have been identified that remove gas from galaxies efficiently, such as ram pressure \citep{gunn72, kenney04, chung09}, harassment \citep{moore96,Moore98,Haynes07}, and strangulation \citep{moore96, bekki98, bekki02}. In addition, feedback from active galactic nuclei (AGN) has been proposed and shown to provide sufficient energy to eject or heat gas, thus preventing the formation of stars \citep{silk98, dimatteo05, fabian12}.

\quad There have been many studies recently exploring how the large-scale structure (LSS) of the universe, also known as the ``cosmic web", affects the formation and development of galaxies. This structure has been observed in galaxy surveys for many years, starting with the iconic CfA redshift survey \citep{delapparent86}, and followed by larger surveys such as the 2-degree Field Galaxy Redshift Survey \citep[2dFGRS,][]{colless01}, the Sloan Digital Sky Survey \citep[SDSS,][]{york00}, the 6-degree Field Galaxy Survey \citep[6dFGS,][]{jones09}, the Galaxy And Mass Assembly Survey \citep[GAMA,][]{driver11}, and the COSMOS Survey \citep[COSMOS,][]{scoville07}. Large simulations with cosmological volumes, including dark matter (DM)-only N-body simulations such as the Millennium simulations \citep{springel05,MBK09}, and hydrodynamical simulations, such as EAGLE \citep{Schaye15} and IllustrisTNG \citep{Nelson18}, have consistently reproduced the pervasive large-scale cosmic web structure in the universe.

\quad The existence of the cosmic web in the formation of structure is well-established. However, it remains unclear how the location of galaxies within this large-scale environment affects their fundamental formation properties, such as mass, star formation, and gas content. Studies have shown that more massive galaxies with redder colors tend to reside closer to the spine of cosmic web filaments, while less massive bluer galaxies are located farther away, at least at low redshift \citep[e.g.,][]{kuutma17,laigle18,luber19}. However, there is a discrepancy in whether filaments enhance or quench star formation. Some research has found that star formation is suppressed near or in filaments \citep[e.g.,][]{kraljic18,Winkel21,Hasan23,Hasan24}, while others have reported the opposite effect \citep[e.g.,][]{Darvish14, Vulcani19,Kotecha22}, although straight comparisons are difficult to make as data and methods are nor directly related. Several studies have investigated the relationship between the cold gas content of galaxies and their location in the cosmic web, but the results have been inconsistent. For example, \citet{odekon18} studied the cosmic web-dependence of {\HI} in galaxies in the ALFALFA survey \citep{giovanelli05} and found that galaxies with masses between $8.5<\log({\ms}/{\Msun})<10.5$ tend to have less {\HI} close to filaments. In contrast, from the MIGHTEE-HI survey \citep{maddox21}, \citet{sinigaglia24} found an excess of {\HI} in $9.5<\log({\ms}/{\Msun})<11.5$ galaxies located in filaments. \citet{kleiner17}, using the HIPASS survey \citep{barnes01}. also reported higher {\HI} fractions near filaments than those further away, but for galaxies with masses $\log({\ms}/{\Msun})\geq11$, 

\quad Cosmological hydrodynamical simulations are excellent tools for testing models of gas accretion from large-scale to galaxies. These simulations are often calibrated to match distributions related to stellar observables, such as the stellar mass/luminosity function \citep[e.g.,][]{Pillepich18} or the star-forming main sequence \citep[e.g.,][]{Schaye15}. However, they are not calibrated for gas observables such as the cold gas fraction. Additionally, all cosmological simulations are limited by resolution and rely on subgrid models of interstellar cold gas and star formation instead of directly modeling them. Comparisons of predictions of gas content in the interstellar medium (ISM), circumgalactic medium (CGM), and intergalactic medium (IGM) with observations allow powerful tests of how well these models reproduce the observed universe \citep[e.g.,][]{Oppenheimer16,Diemer19,Stevens19,Hasan20}. 

\quad Recently, \citet{Dave20} compared cold gas in various simulations with different subgrid prescriptions and numerical methods. They showed that while statistical distributions such as the {\HI} mass function and gas scaling relations with stellar mass and star formation rate are qualitatively reproduced across different simulations (in the local universe), there are quantitative differences with galaxy properties. For example, SIMBA \citep{Dave19} and IllustrisTNG \citep{Diemer19} produce an excessive amount of {\HI} relative to observations in massive galaxies, whereas EAGLE \citep{Crain17} produces insufficient {\HI} in green valley galaxies. The amount of {\HI} produced in different simulations is significantly influenced by feedback from internal processes such as black holes. Additionally, numerical differences in resolution also have an impact on the amount of measured {\HI}. For example, the {\HI} content of the CGM and the IGM depend strongly on resolution and do not converge \citep[e.g.,][]{Hummels19,Peeples19,Mandelker21}. In a handful of simulation studies that explore the filamentary dependence of gas in galaxies, some find a depletion of gas close to the LSS \citep[e.g.,][]{BL13,zhu22,Hasan23,Bulichi24}, while others report higher gas fractions close to the LSS \cite[e.g.,][]{Singh20,RG22}.

\quad The relationship between the gas and star formation properties of galaxies and their cosmic web environment has led to different interpretations of the connection between galaxies and LSS. According to the cold-accretion model of galaxy formation, narrow filaments or streams supply cold gas to galaxies, thereby causing them to grow at higher redshifts \citep[e.g.,][although some argue for shock-heating of gas to first form a thin stellar disk, e.g., \citealt{Stern21,afruni23}]{Dekel06,Dekel09,Pichon11}. Building on this idea, \citet{cwd} proposed the idea of ``cosmic web detachment'', wherein galaxies are cut off from narrow cold gas-supplying filaments at lower redshift due to mergers, satellite interactions, etc., leading to their gas supplies being used up and quenching star formation as a result. Furthermore, several works have shown that hydrodynamical instabilities can disrupt cold streams in massive halos with a hot CGM if gas cooling times are long \citep[e.g.,][]{Mandelker20a,Mandelker20b,Daddi22a,Daddi22b}. Filaments and LSS have been linked to quenching through other gas removal mechanisms, such as ram stripping \citep[e.g.,][]{Zinger18} or accretion shock heating \citep[e.g.,][]{Pasha23}. Conversely, an alternative framework proposes that the cosmic web increases gas accretion and subsequent star formation in galaxies \citep[e.g.,][]{Birnboim16}. To establish a more accurate physical framework for understanding the effect of the LSS on galaxy formation, a consistent comparison between simulation predictions and empirical data is crucial.

\quad In this work, we use neutral hydrogen (HI) data from the COSMOS {\HI} Large Extragalactic Survey (CHILES, van Gorkom et al. in prep) to investigate the impact of the LSS on galaxy evolution as a function of redshift. CHILES is a 1027-hour survey of a single pointing of the Karl G. Jansky Very Large Array (VLA) coincident with the COSMOS field in the VLA-B configuration, the second most extended configuration. The full CHILES survey can detect and resolve {\HI} emission from the most {\HI} massive galaxies continuously out to a redshift of 0.48. In addition, the nature of CHILES being a deep blind survey well positions it for usage in stacking experiments (Lubet et al. submitted). By analyzing the data, we can determine the redshift evolution of the average gas content of galaxies in different LSS environments, in a continuous redshift range from 0.1 $< z <$ 0.48. We also compare the dependence of galaxy {\HI} content and star formation activity on the LSS observed through CHILES with that predicted by the IllustrisTNG cosmological simulations.

\quad This paper is structured as follows. In Section \ref{sec:data}, we discuss the observational data, describing the CHILES data and the ancillary data in COSMOS, and the data used from the IllustrisTNG simulation. Here we also outline the procedure for observational stacking techniques and the method by which we identify the large-scale structure of the universe and the distance of a galaxy to these structures. In Section \ref{sec:results}, we define the galaxy properties and redshift ranges over which we take our measurements, and in Section \ref{sec:disc}, we discuss our results in the context of current observational and theoretical work. Finally, in Section \ref{sec:conc}, we summarize our comparisons. In this work, we assume the {\it Planck 2015} cosmology \citep{Planck15}, with $H_{0}=67.74$ {\kmsmpc}, $\Omega_{\mathrm{M},0} = 0.3089$, and $\Omega_{\Lambda,0} = 0.6911$. All distances are quoted in comoving units, unless stated otherwise.

\section{Data and Methodology} \label{sec:data}

\subsection{CHILES Data}\label{sec:chiles_data}

\quad The CHILES data used in this work have been calibrated and flagged by the automated pipeline presented in Pisano et al. (in prep). The data were then continuum subtracted, imaged, and flagged using the procedure presented in \citet{LuberStacking}. Of the 1027 hours that were observed as a part of the CHILES project, the data used in this work amounts to 802 hours of observation. This difference arises from the need for more detailed calibration and/or imaging for the remaining 200 hours due to the presence of strong radio frequency interference in these databases. However, not including these data only translates to approximately 13\% greater noise. The {\HI} cube used in this work has also been smoothed to a common resolution of 9$\arcsec$ for the entire bandwidth, which spans from 960 - 1420 MHz, with a channel width of 125 kHz, and an image size of 40$\arcmin$. In this work, we limit our study to the redshift range 0.09 $< z <$ 0.48 where we achieve spatial resolution of 15.1 - 53.7 kpc, a velocity resolution of 31.4 - 57.7 km s$^{-1}$, and field of view of 4.03 - 14.33 Mpc, at $z=0.09$ and $z=0.48$, respectively. Given these parameters, even at the highest redshifts most galaxies with stellar mass above 10$^{9}$ M$_{\odot}$ will be kinematically resolved, and the largest galaxies, stellar mass above 10$^{10}$ M$_{\odot}$, will be slightly spatially resolved as well. 

\subsection{Multiwavelength Observational Data}\label{sec:mw_data}

\quad Throughout this work, we use the extensive multi-wavelength data and subsequent derived galaxy properties, made publicly available by the COSMOS collaboration \citep{scoville07}. From the public COSMOS data, we use the COSMOS2008 ID and the rest frame NUV-r color, and derived stellar masses, presented in \citet{laigle16}. In addition to the COSMOS data, we also use data made available by the Galaxy and Mass Assembly (GAMA) survey \citep{driver11}. Specifically, we use the GAMA collaboration's reprocessing of COSMOS multi-wavelength data to derive the highest confidence redshifts \citep{davies15}. For the {\HI} stacking done in this work, it is critical to have a-priori systemic redshifts with errors an order of magnitude less than the typical {\HI} line width in the stack, to ensure that we can correctly identify the expected line center of the {\HI} detection and not smear the average spectrum \citep{maddox13}.  As a result, in all input galaxies for the {\HI} stacks in this paper, we only used galaxies with the highest confidence redshifts, as specified in \citet{davies15}.

\subsection{TNG Simulation data}\label{sec:tngdata}

\quad From the simulation side, we employ the IllustrisTNG magneto-hydrodynamical cosmological simulations \citep{Pillepich18,Nelson18,TNGDR19,Springel18,Naiman18,Marinacci18}, which simulated the evolution of gas, stars, DM, and black holes (BH) from the early universe ($z=127$) to the present day ($z=0$) using the AREPO moving-mesh hydrodynamics code \citep{Springel10}. In particular, we make use of TNG100-1, the highest resolution run of the TNG100 simulation, which has a box size of $\sim$110.7 comoving Mpc per side, a minimum baryonic and DM particle mass of $\sim\!1.4 \!\times\! 10^{6}~{\Msun}$ and $\sim7.5 \!\times\! 10^{6}~{\Msun}$ respectively, and $1820^3$ initial DM particles. 

\begin{figure}[ht!]
\begin{center}
\includegraphics[width=\columnwidth]{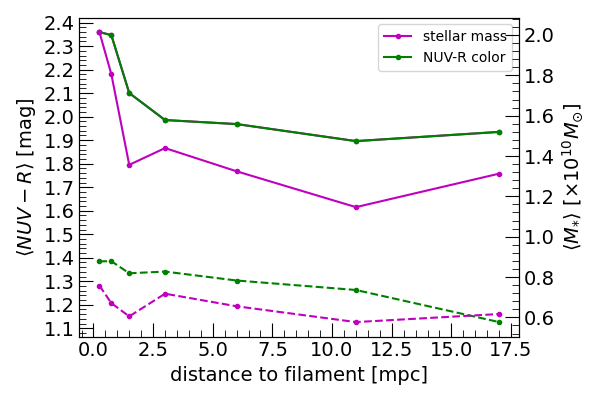}
\caption{The average stellar mass (green) and NUV-r color (magenta) for the COSMOS field for all galaxies (solid lines), and for galaxies in our stellar mass and color criterion (dashed lines).}\label{fig:ColorMass}
\end{center}
\end{figure}

\quad We obtain galaxy data for all 100 snapshots of the TNG100-1 simulation (hereafter TNG) from the online data repository\footnote{\url{https://www.tng-project.org/data/}} \citep{TNGDR19}. To identify dark matter (sub)halos in the simulation, a Friends-of-Friends (FoF) algorithm with a linking length = 0.2 is first applied to the DM particles. Gas and stars are then assigned to FoF groups based on their nearest-neighbor DM particle. Finally, the {\sc Subfind} algorithm \citep{Springel01,Dolag09} is applied to the total mass distribution in each FoF group. The most massive {\sc Subfind} object in each FoF group is identified as the central halo. For each snapshot, we identify all galaxies with a minimum stellar mass of $\log({\ms}/{\Msun}) \!=\! 9$ for consistent comparison with the observed data. The galaxies in our sample are, therefore, well resolved with $\gtrsim$700 stellar particles in them. The catalogs contained $\sim$15,000--20,000 galaxies for each snapshot at $0 \leq z \leq 0.5$. 

\quad For each of the snapshots, we obtained the following quantities for each galaxy: galaxy comoving position, star formation rate (SFR), stellar mass ({\ms}), halo virial radius ({\Rh}; comoving radius enclosing an average overdensity of 200 times the critical density of the Universe), halo mass ({\mh}; the mass within {\Rh}), mass of all gas gravitationally bound to a subhalo ({\mgas}), mass of all gas within twice the stellar half-mass radius ({\mgasre}), and stellar photometric magnitude in the Johnson filter U-band \citep{Buser78} and SDSS $r$-band \citep{Stoughton02}. The synthetic colors are dust-corrected as in \citet{Nelson18}.

\quad We also obtained the {\sc HI} gas mass of galaxies at redshifts $z=0$, 0.5, 1, 1.5, and 2 using the post-processing framework to estimate the abundance of atomic and molecular hydrogen developed by \citet{Diemer18}. They tested five different models for the atomic-to-molecular transition, based on empirical correlations, high-resolution simulations of isolated galaxies, and analytic modeling. The catalogs they produced of {\HI} gas mass and density profiles of galaxies are available on the TNG data repository. These catalogs were compared with the observations in \citet{Diemer19}. Here, we use just the {\HI} gas masses based on mock 2D projections on the sky within a radial distance of 60 kpc from a galaxy, in order to fairly compare the observational results to the simulations. We consider the different models of {\HI}/{\Hmol} to assign a {\HI} mass, {\MHI} to each galaxy, as well as the $\pm1\sigma$ uncertainty in this value. We also obtained SFRs from 2D SFR surface density maps presented in these catalogs. These measurements are described in more detail in Section~\ref{sec:tngmeasure} below.

\subsection{Defining the Cosmic Web}\label{sec:cw_def}

\quad Here, we describe our procedure for reconstructing the cosmic web from galaxy catalogs in both simulations and observations and for defining the distance of a galaxy to the nearest filament of the cosmic web.

\quad To investigate the effects of the large-scale structure on the universe, several mathematical frameworks and algorithms are used to identify the filaments that make up the cosmic web (see \citet{libeskind18} for a review of different techniques). For this work, we use the Discrete Persistent Source Extractor (DisPerSE), which is a scale-free topological algorithm that uses discrete Morse theory to compute the gradients in topological features and return the filamentary structure \citep{sousbie11a, sousbie11b}. The filamentary structure is defined as a system of critical points that can be connected to form the resulting composite structure. DisPerSE has been used to identify LSS for several {\HI} studies \citep{odekon18, kleiner17, tudorache22, Hoosain24} and studies using the COSMOS dataset \citep{kleiner17}. It has also been used to identify filaments in TNG \citep{GE22,Malavasi22,Hasan23,Hasan24}. Below, we give a brief description of how {\disperse} works.

\quad {\disperse} first takes a set of inputs, i.e., the positions of galaxies, and uses it to segment the volume into tetrahedrons, whose vertices are the galaxy locations. This technique, referred to as the Delaunay Tessellation Field Estimator \citep[DTFE; ][]{SV00,VS09}, constructs a density field where the density values at the galaxy locations are measured directly, and the density is linearly interpolated at other spatial locations. Next, the gradient of the density field is computed, and discrete critical points are identified where the gradient vanishes, including minima (corresponding to voids), maxima (corresponding to nodes), and saddle points. Filaments, made up of individual segments, follow the ridge-lines of the density field and connect maxima to saddle points. In practice, a persistence parameter is set in order to define a robustness threshold with respect to Poisson noise (analogous to a signal-to-noise ratio), and this outputs structures such as critical points and filaments with the corresponding significance.

\subsubsection{The Observed Cosmic Web}

\quad In \citet{luber19}, we use DisPerSE specifically for the CHILES field within COSMOS to investigate the expected neutral hydrogen properties of galaxies as a function of placement in the large-scale structure. Additionally, \citet{hess19} and \citet{bluebird20} identified the LSS using {\disperse}, then compared the {\HI} properties of galaxies in groups and the {\HI} spin vectors for nearby galaxies, respectively, as a function of their placement in the LSS. In this work, we applied DisPerSE to the entire COSMOS field using GAMA spectroscopic redshifts (as described in Section \ref{sec:mw_data}) using the mirror boundary condition and a critical point significance level of 4 (see \citet{sousbie11a} for a description of these parameters and \citet{luber19} for examples of them applied to CHILES data). This implementation differs from \citet{luber19} as it is done in three dimensions, as opposed to two dimensions, allowing us to properly recover the physical distances from a galaxy to the closest cosmic web spine. The input to {\disperse} was the positions of all galaxies in this catalog with a minimum stellar mass ${\logms}\geq9$, corresponding to the approximate mass-completeness limit of COSMOS.

\begin{figure*}
\begin{center}
\includegraphics[width=\textwidth]{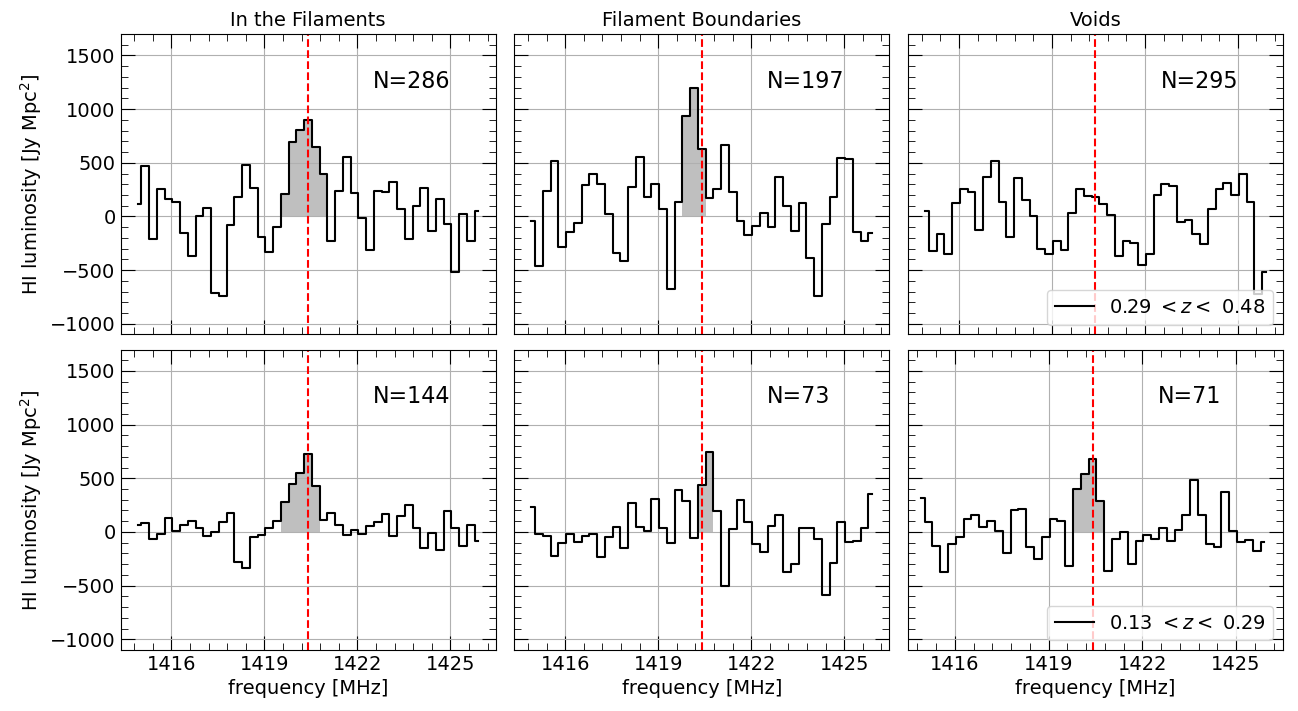}
\caption{The stacked spectra using CHILES, in units of {\HI} line luminosity, for the high redshift bin (top; $0.29<z<0.48$), and the low redshift bin (bottom; $0.13<z<0.29$) for galaxies in the 0 $< D_{fil}/\mathrm{Mpc} <$ 2 (left; corresponding to filament cores), 2 $< D_{fil}/\mathrm{Mpc} <$ 4 (middle; filament boundaries), and 4 $< D_{fil}/\mathrm{Mpc} <$ 20 (right; voids) bins. The gray-shaded region in each spectrum corresponds to the channels integrated to calculate the average {\HI} masses and the red dashed line indicates the rest frequency. The number in the top right of each spectrum corresponds to the number of input galaxies for the stack.}
\label{fig:StackSpectra}
\end{center}
\end{figure*}

\subsubsection{The Simulated Cosmic Web}

The reconstruction of cosmic web structure from galaxy catalogs in TNG is similar to the procedure in \citet{Hasan23}. In this work, we apply the {\disperse} algorithm to galaxy catalogs with a minimum stellar mass of ${\logms}=9$ at each snapshot for a construction comparable to the observed catalogs. We choose a persistence threshold of 4$\sigma$ as in the COSMOS reconstruction but also vary this to be 3$\sigma$ and 5$\sigma$. Our resulting statistics -- including sSFR and {\HI} gas fraction -- are not very sensitive to persistence, changing at most by 5\% between 3$\sigma$ and 5$\sigma$ persistence. We also smoothed the position of the segments of the filamentary skeleton to reduce sharp/unphysical shapes of segments caused by shot noise. Here, the initial positions of the extrema of a local filament segment are averaged with those of the contiguous segments, thus amounting to a length scale of $<$1~Mpc (since filament segments are comparable to this length scale).

\subsubsection{Defining Distances}

\quad Determining the accurate distance of a galaxy from the nearest filament is crucial for this study. However, it can be somewhat complicated, as the output from DisPerSE is not a smooth analytic definition, but rather a series of critical points that comprise a filament. To overcome this, we have devised the following method for calculating the distances from the filaments. First, we identify the critical point of the filamentary structure that is closest to the galaxy. Next, we locate the critical point that is the second closest to the galaxy, but on the same filament as the first closest critical point. If the galaxy lies between these two points, we draw a straight line between them and calculate the length of a line drawn from the galaxy to the line, forming a right angle. If the galaxy is not between two points and resides beyond the line connecting the filament, we return the distance to the nearest critical point as the distance from the filament to the galaxy.

\quad In the TNG simulations, we calculate the transverse distance of a galaxy from the closest identified filament spine as in \citet{Hasan23}. We define the distance from a galaxy to the nearest segment midpoint to be {\dfil}. Although this definition differs slightly from the {\dfil} calculated for the CHILES galaxies, it is effectively equivalent. The segments in the TNG simulation are composed of critical points always less than 1 Mpc apart and hardly ever more than 0.5 Mpc apart. Additionally, the smoothing of the filamentary skeletons typically smooths structures on scales of around 1-2 Mpc/h \citep{Cohn22}. We verified that there was no noticeable change in the simulation statistics (including effect of distance on galactic properties) whether we used the \citet{Hasan23} definition involving segment midpoints or the \citet{luber19} definition involving critical points. Moreover, the defining of large-scale structure yields a distribution of galaxies, as a function of filament distance, for both the simulations and observations of similar slope and 50\% anchoring point.

\begin{deluxetable*}{ccccccc}[t!]
\tablecaption{Results of {\HI} emission stacking in two redshift bins and three bins of {\dfil} in CHILES.\label{tab:obs}}
\tablecolumns{7}
\tablenum{1}
\tablewidth{0pt}
\tablehead
{
\colhead{Redshift} &
\colhead{D$_{fil}$} &
\colhead{N$_{gal}$} &
\colhead{Average {\HI} Mass} &
\colhead{Average S.F.R.} &
\colhead{Average Stellar Mass} &
\colhead{Significance\tablenotemark{a}} \\
\colhead{} & 
\colhead{Mpc} &
\colhead{} &
\colhead{10$^{9}$ M$_{\odot}$} &
\colhead{M$_{\odot}$ yr$^{-1}$} &
\colhead{10$^{9}$ M$_{\odot}$} &
\colhead{} 
}
\startdata
0.13 - 0.29 & 0 - 2  & 144 & 2.0$\pm$0.4 & 0.23$\pm$0.02 & 8.6  & 5.5 \\
0.13 - 0.29 & 2 - 4  & 73  & 1.1$\pm$0.4 & 0.22$\pm$0.03 & 6.4  & 3.9 \\
0.13 - 0.29 & 4 - 20 & 71  & 1.2$\pm$0.5 & 0.20$\pm$0.01 & 8.0  & 3.8 \\
0.30 - 0.48 & 0 - 2  & 286 & 2.7$\pm$1.0 & 0.31$\pm$0.03 & 9.0  & 3.0 \\
0.30 - 0.48 & 2 - 4  & 197 & 2.3$\pm$0.8 & 0.42$\pm$0.02 & 10.0 & 3.6 \\
0.30 - 0.48 & 4 - 20 & 295 & $<1.2$\tablenotemark{b}      & 0.29$\pm$0.03 & 9.1  &  -  \\
\enddata
\tablenotetext{a}{We define the significance of the {\HI} stacked detection by dividing the peak flux in the spectra by the r.m.s. noise of the spectra, as measured in the line-free channels. These spectra correspond to the integrated {\HI} profiles shown in Figure \ref{fig:StackSpectra}.}
\tablenotetext{b}{This value corresponds to the 3$\sigma$ detection limit for this non-detection.}
\vspace{-10pt}
\end{deluxetable*}
\vspace{-25pt}

\section{Results}\label{sec:results}

\quad In this study, we analyze the redshift evolution of the {\HI} gas fraction and the specific star formation rate for blue galaxies in both simulations and observations in different large-scale environments. For the observational measurements, we define blue galaxies as those with a rest-frame NUV-r color in the range -1 to 3. This is in order to maximize the {\HI} detection in our stacked spectra, following authors such as \citet{catinella10} who report very low ($<$10\%) {\HI} gas fractions for nearby galaxies with NUV-r$>3$, and \citet{LuberStacking} where we find no detectable HI for red galaxies in the CHILES field at any redshift. Since TNG does not provide GALEX NUV magnitudes, for these measurements, we define blue galaxies as U-r$\leq1$, based on the conversion from AB to Vega magnitudes and NUV-r to U-r color \citep[e.g.,][]{laigle16}. We find that this color cut does indeed result in a bimodality in color-magnitude space at $z\leq0.5$ in TNG, by separating red and blue galaxies. We verify that a small change ($\sim$10\%) in the color cut does not affect our conclusions. These criteria ensure that we are stacking galaxies that lie in the blue cloud of color-magnitude space. In Figure \ref{fig:ColorMass}, we show the NUV-r color and stellar mass for all galaxies in the COSMOS field (solid lines), and for galaxies that remain after our color and mass selection (dashed lines). Here we see that when including all galaxies, we reproduce the well understood trend that galaxies become redder and more massive closer to filaments. However, the galaxies within our selection criterion show no change in environment, as a result of our criterion. Thus, we probe the question of how the environment affects the {\HI} and SFR of like galaxies in different environments.

\quad To properly probe the redshift evolution of blue galaxies, we must define both redshift bins and large-scale structure bins. For our large-scale structure bins, we require at least one bin to be centered on a galaxy-cosmic web separation of 3.03 Mpc, the median $D_{fil}$ of CHILES galaxies, with other bins representing a range of cosmic web environments. As a result, we choose the following bins. $D_{fil}$=0--2~Mpc, which is representative of galaxies most closely associated with filaments, $D_{fil}$=2--4~Mpc, which is representative of galaxies in a transitional density field between filament environment and non-filament environment, and $D_{fil}$=4--20~Mpc, which represent the galaxies residing in voids.

\quad The continuous redshift coverage of CHILES allows us to construct contiguous redshift bins in the redshift range we are choosing to investigate. We limit our study to redshifts in the range 0.13 $< z <$ 0.48. We choose 0.13 as our lower bound, as it is the lowest range in which we believe the physical spatial scale ($\approx6$~Mpc) is large enough to properly sample the cosmic web, and 0.48 as our upper bound as it is the upper bound of the CHILES observations. Additionally, we split our redshift range into two ranges, 0.13 $< z <$ 0.29 and 0.30 $< z <$ 0.48. Choosing these two ranges results in approximate uniform sensitivity for our stacks, as the lower redshift bin has less galaxies but greater {\HI} mass sensitivity, and the higher redshift bin has more galaxies and lesser {\HI} mass sensitivity. The average HI mass sensitivity is $ M_{HI}\sim10^{9}~{\Msun}$ across the two redshift bins, and the stellar masses are complete to the level of ${\ms}\sim10^{9}~{\Msun}$ \citep{Ilbert09}.

\quad With these selections, we have defined 6 bins in which we conduct stacks in order to probe the cosmic-web dependent evolution of {\HI} gas fraction. Below, we summarize our methodology for measuring the galaxy properties with this setup for both the observations and the simulations and compare the average measurements in these sub-samples.

\subsection{{\HI} and Star Formation Measurements from Observations} \label{sec:obsmeasure}

\quad We measure the average {\HI} masses for our different samples using the {\HI} cubelet stacking technique. This technique involves extracting a small cubelet that encompasses each galaxy, averaging all the cubelets together, and CLEANing the stacked emission \citep[see][for the description and capabilities of the cublet stacking technique, and \citet{LuberStacking} for its adaptation for the CHILES data]{chen21a,chen21b}. The resulting spectra from our six {\HI} stacks are shown in Figure \ref{fig:StackSpectra}, where from left-to-right we see the three increasing D$_{fil}$ bins for both the high redshift (top row) and low redshift (bottom row) ranges we probe. For the spectra, we integrate the {\HI} emission over the equivalent of a 60~kpc box at the average redshift for each bin. The choice of 60 kpc is to ensure that we include the majority of all detectable {\HI} emission while also not significantly adding to the noise of the spectra. Each panel shows emission channels identified by the grey-shaded region. We identified these channels by first finding the positive and negative velocity channels that cross the expected one-sigma value relative to the rest-frame line center and then including all channels that lie between them.
We used these channels to calculate the average {\HI} mass. The errors on the {\HI} mass are calculated by taking the integrated flux over 1000 regions in the stacked {\HI} cube and converting the r.m.s. of the distribution of fluxes into an error in {\HI} mass.

\quad We implemented a similar technique to measure the average star-formation rate. Instead of cubelets, we stack small 1.4 GHz radio continuum images centered on each source. For each of these images, we scale them to 1.4 GHz by using the redshift for each source and assuming that the radio continuum intensity follows a power law relationship with a spectral index of -0.75. These images are then averaged together, with the same weights as the {\HI} stack, we average flux density in the stacked image, and then convert it into a star-formation rate using the scaling relationship presented in \citet{murphy11}. The errors of this measurement are simply the r.m.s. noise of the stack image converted into star-formation rate. However, we note that this does not account for any increase in signal due to confusion, AGN contamination, and intrinsic uncertainty in using radio continuum intensities to measure star-formation rate. The former two could result in our measure of star-formation rate being too high, while the latter may result in a systematic bias in either direction. We measure the average stellar mass simply by averaging the stellar masses for each sample, as reported in \citet{laigle16}, using the same noise-based weights used in the average {\HI} mass and SFR calculations. For explicit details, we refer the readers to \citet{LuberStacking} for a thorough description of the stacking technique, and any limitations that the methods may have.

\quad As a summary of our observations, we present all observationally measured parameters, including average {\HI} mass, star-formation rate, and stellar mass, in Table~\ref{tab:obs}. Additionally, the last column of Table~\ref{tab:obs} corresponds to the significance of the {\HI} detection. We calculate this by dividing the peak {\HI} flux by the r.m.s. noise in the rest frequency range 1415 - 1418.5 MHz and 1422.5 - 1426 MHz. With this statistical measure, we find that the stack for the high-redshift bin for the void galaxies did not yield a significant {\HI} detection to the 3$\sigma$ limit, and we report a 3$\sigma$ upper limit for the corresponding {\HI} mass, and gas fraction. 

\begin{figure*}
\begin{center}
\includegraphics[width=\textwidth]{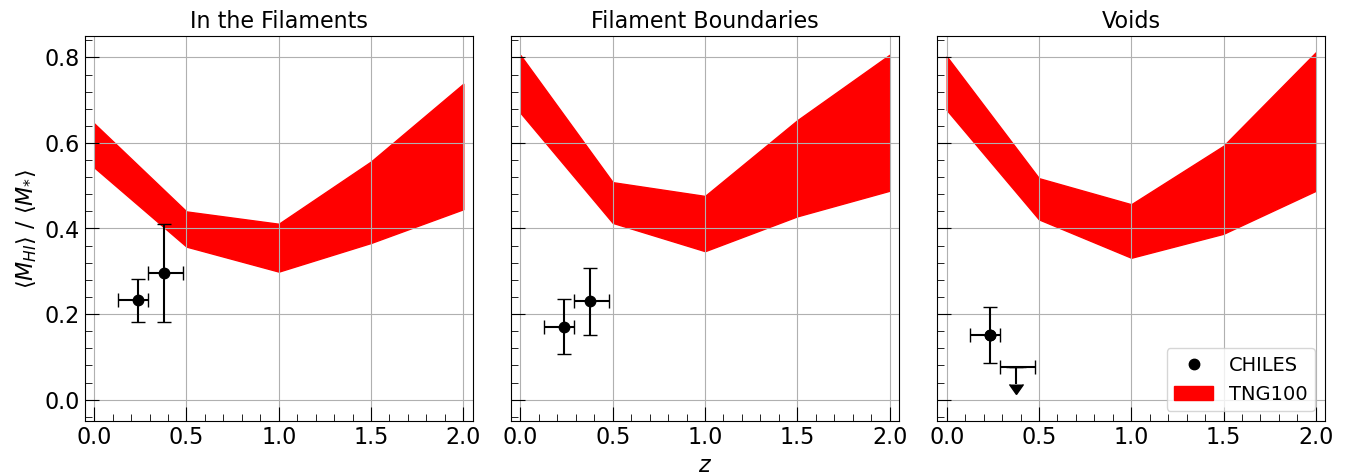}
\caption{The mean {\HI} gas fraction derived from the CHILES data (black markers) and the TNG data (red shaded region) as a function of redshift for three different cosmic web environments, specifically, the distance to the filament bins of 0 $< D_{fil}/\mathrm{Mpc} <$ 2 (left), 2 $< D_{fil}/\mathrm{Mpc} <$ 4 (middle), and 4 $< D_{fil}/\mathrm{Mpc} <$ 20 (right) bins.
The simulations overproduce the {\HI} fraction relative to observations. The direction of redshift evolution is also in disagreement in filament cores and boundaries, but agrees in the voids.}
\label{fig:fHI_z}
\end{center}
\end{figure*}

\begin{figure}
\begin{center}
\includegraphics[width=\columnwidth]{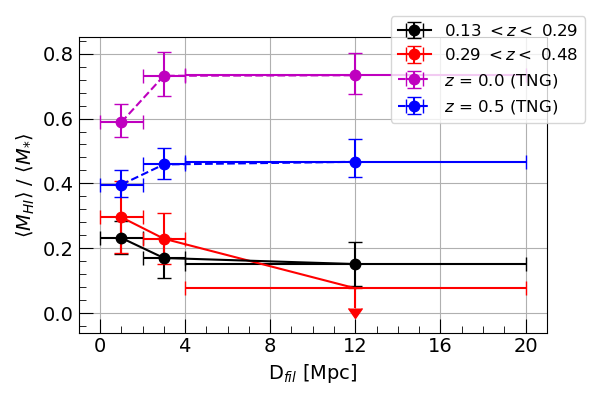}
\caption{The mean {\HI} gas fraction as a function of distance to the filament in two redshift bins for CHILES data (red is higher, black is lower) and TNG simulations (blue is higher, purple is lower). CHILES shows a decrease away from filaments at low-$z$, the opposite of what TNG predicts.}
\label{fig:fHI_dfil}
\end{center}
\end{figure}

\begin{figure*}
\begin{center}
\includegraphics[width=\textwidth]{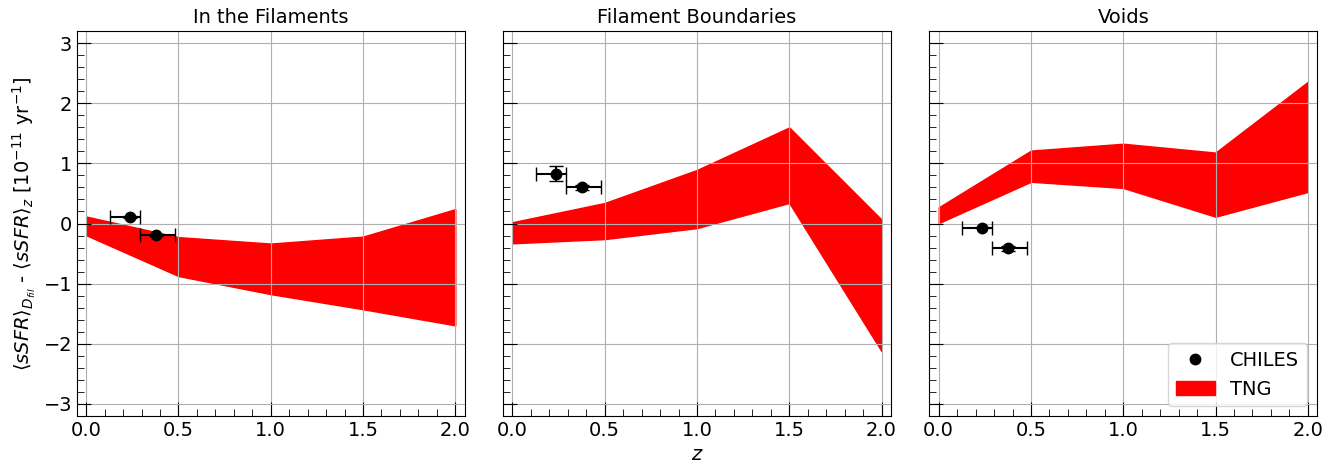}
\caption{The deviation from mean sSFR derived from the CHILES data (black markers) and the TNG data (red shaded region) as a function of redshift for three different cosmic web environments as we define in Figures ~\ref{fig:StackSpectra} and \ref{fig:fHI_dfil}. In both redshift ranges, the strongest tension is in the voids, where CHILES finds the least star formation, and TNG predicts the most star formation. There is significantly better agreement in the denser regions, especially the filament boundaries at $z\sim0.1-0.3$.}
\label{fig:ssfr_z}
\end{center}
\end{figure*}

\begin{figure}
\begin{center}
\includegraphics[width=\columnwidth]{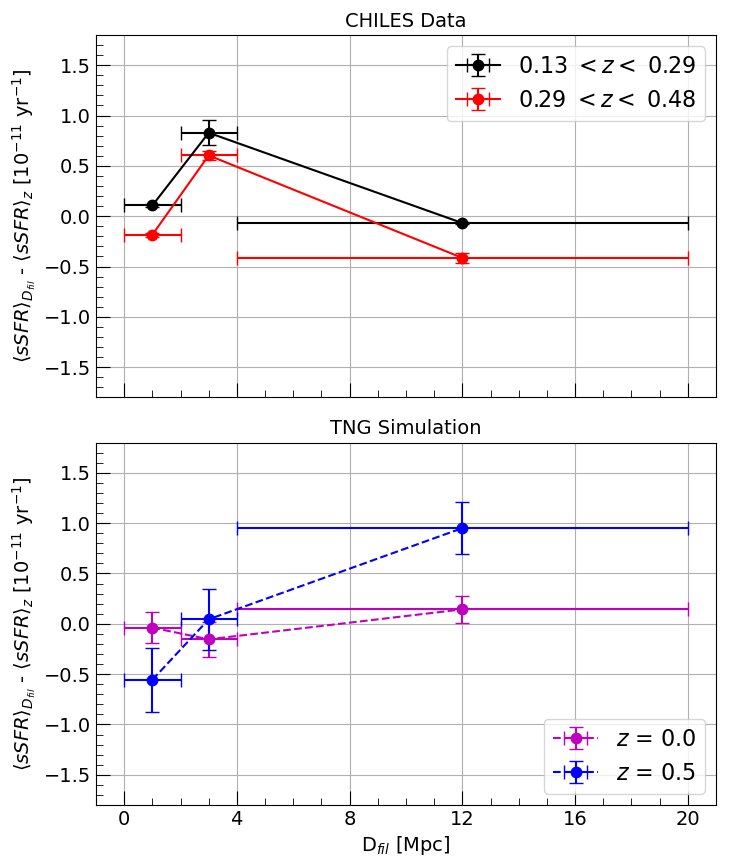}
\caption{The deviation from mean sSFR as a function of distance to the filament in two redshift bins in CHILES (top) and TNG (bottom).
Unlike in CHILES, TNG generally predicts increased star formation with distance to filaments (more so at $z=0.5$).}
\label{fig:ssfr_dfil}
\end{center}
\end{figure}

\subsection{{\HI} and Star Formation Measurements in TNG} \label{sec:tngmeasure}

\quad In the TNG simulations, each gas cell has a neutral hydrogen fraction that represents the ratio of hydrogen mass to total gas mass. The hydrodynamics and star formation of galaxies in TNG are determined by the two-phase subgrid ISM model of \citet{SH03}, which assumes that star formation occurs in cells above a gas density threshold of $n_{\mathrm{H}} \simeq 0.1~\mathrm{cm}^{-3}$ and that the SFR is proportional to the fraction of cold gas. In star-forming cells, the neutral hydrogen fraction is taken as simply the cold gas fraction, which is between 0.9 and 1, with a minor dependence on density. The computation of neutral fraction in non-star forming cells is a lot more involved and accounts for various sources of cooling, including the UVB, and self-shielding of gas \citep[see, e.g.,][]{Vogelsberger13}. As a result, the neutral hydrogen fraction in TNG has some uncertainty associated with it (see section~\ref{sec:disagree}).

\quad \citet{Diemer18} post-processed the TNG100 data to estimate the {\HI}/{\Hmol} abundance of each galaxy by developing an improved treatment for UVB and comparing the neutral hydrogen fraction estimated from 2D face-on maps of a galaxy projected on the plane of the sky with that estimated for each cell in 3D (a ``volumetric'' model). While the population-averaged predictions for the neutral fraction agreed reasonably well, there is significant variance in the neutral fraction of individual galaxies between the 2D projected and 3D volumetric models. We used the projected neutral fraction for three reasons: (1) the projected neutral fraction is more comparable to the 2D {\HI} maps obtained in the observations; (2) the projected modeling is less sensitive to numerical resolution than volumetric modeling; (3) \citet{Diemer18} showed that volumetric modeling is biased and leads to large cell-to-cell scatter in surface densities when using the Jeans length approximation to estimate the column densities of gas cells. However, we did not find a large variation (more than a few percent) in the mean {\HI} fraction between the projected and volumetric models.

\quad \citet{Diemer18} examined five post-processing models that calculate the {\HI}/{\Hmol} transition. These models were based on observed correlations \citep{Leroy08}, idealized analytical models of gas clouds \citep{Krumholz13,Sternberg14}, and full high-resolution simulations of isolated galaxies \citep{GK11,GD14}. Although the models generally agreed, the neutral fraction and spatial distribution of {\Hmol} in individual galaxies differed according to specific modeling details. Furthermore, \citet{Diemer19} found that TNG reasonably reproduced the median and scatter in the observed {\HI} and {\Hmol} fractions at $z=0$. However, there was a significantly larger fraction of satellite galaxies with very little to no neutral gas in TNG compared to the observed universe. They attributed this phenomenon to strong environmental processes from large DM halos, which efficiently strip the gas from these galaxies, as well as excessive feedback blowing the gas away.

\quad For each galaxy, we calculate the 2D projected {\MHI} within an aperture of 60~kpc from the galactic center of each galaxy from the five {\HI}/{\Hmol} models described above. This aperture size is chosen to match the integration area of the stacked {\HI} detection in CHILES. To each galaxy, we assign a nominal value of {\MHI} as the median {\MHI} of the models, and the $-1\sigma$ and $+1\sigma$ {\MHI} values as the \nth{16} and \nth{84} percentile values of the models, respectively. We used these values instead of the arithmetic mean, minimum, and maximum of the model outputs, as those might bias the estimates to outliers. We calculate the nominal {\HI} gas fraction, ${\fHI} = {\MHI}/{\ms}$, for each galaxy at redshifts $z=0$, 0.5, 1, 1.5, and 2. The nominal {\MHI} values are used to calculate the mean {\HI} fraction, {\meanfHI}, as presented below, while the $\pm1\sigma$ {\MHI} values are incorporated into the calculation of uncertainties in {\meanfHI}.

\quad The specific star formation rate (sSFR) we obtained represents the 2D projected values from the catalogs provided by \citet{Diemer18}. Each gas cell has an associated hydrogen mass and an instantaneous SFR. Using the subhalo catalogs, we sum up these measurements for all particles and cells that are gravitationally bound to a particular subhalo. We then compute the sSFR by dividing the instantaneous SFR of all gas cells within an aperture of 60 kpc from galaxy center -- to be consistent with the observational estimation -- projected onto the two-dimensional plane of the sky by the total stellar mass ({\ms}). 

\quad We analyzed the mean sSFR and {\fHI} in the IllustrisTNG simulation for galaxies with stellar mass ${9 \leq \log({\ms}/{\Msun}) \leq 10.5}$ to compare with observations. We divided the galaxies into three bins of distance to filament spines ({\dfil}): ${\dfil}=0-2$~Mpc (filament cores), ${\dfil}=2-4$~Mpc (filament outskirts), and ${\dfil}=4-20$~Mpc (voids). We then computed the mean deviation from the average sSFR ({\meanssfr}) and the mean {\fHI} (defined as ${\meanfHI} = {\meanMHI}/{\meanMs}$) for each radial bin at redshifts $z=0$, $0.5$, $1$, $1.5$, and $2$. We determined the $\pm1\sigma$ uncertainties by adding in quadrature the bootstrapped errors on the values of {\meanfHI} and the $\pm1\sigma$ values of {\meanfHI} from the models, as described in Section~\ref{sec:tngmeasure}. We choose mean deviation from sSFR {\meanssfr} instead of mean sSFR in order to understand which cosmic web environments have more or less star formation than the average level at a given redshift, i.e., {\it relative} star formation activity, rather than an {\it absolute} measure of star formation activity.

\subsection{Stacking Results}\label{sec:StackingRresults}

\quad Here, we compare simulations and observations of the {\HI} gas fraction and deviation from mean sSFR across the redshift and cosmic web environment using the data, techniques, and galaxy subdivisions outlined above.

\subsubsection{{\HI} Gas Fraction}\label{sec:fhi_results}

\quad Using the procedures defined above, we summarize our result for the mean {\HI} gas fraction in Figures \ref{fig:fHI_z} and \ref{fig:fHI_dfil}. Figure~\ref{fig:fHI_z} presents the average {\HI} gas fraction as a function of redshift for the observations (black markers) and simulation (red shaded region) across three different cosmic web environments. Similarly, Figure~\ref{fig:fHI_dfil} shows the mean {\HI} gas fraction as a function of D$_{fil}$, separated by redshift bin.

\quad The simulated and observational data show significant disagreements.
TNG always overpredicts the observed mean {\HI} fraction in each cosmic web environment and redshift bin, in some cases by a factor of $\gtrsim$2. In the filament cores and boundaries, the redshift evolution is also different between TNG and CHILES (Figure~\ref{fig:fHI_z}). TNG predicts a strong decline from $z=0$ to 0.5, while the CHILES observes almost no evolution within the errors. However, for the voids, where the simulated {\HI} fraction is at least 3 times larger than observed, both TNG and CHILES show a declining {\meanfHI} from lower to higher redshift. However, the observational measurement at the higher redshift in the voids has the weakest signal and is an upper limit. Studying the mean {\HI} fraction as a function of distance to the filaments, we find that the theoretical and observational trends are reversed here too (Figure~\ref{fig:fHI_dfil}). In particular, {\meanfHI} decreases from filament cores and outskirts to voids at a fixed redshift bin in CHILES, while TNG predicts an increase in {\meanfHI} further away from the filaments. In CHILES, the drop in {\HI} fraction with filament distance is much stronger in the higher-redshift ($0.3<z<0.48$) bin. In TNG, the rise in {\HI} fraction with filament distance is stronger in the lower-redshift snapshot ($z=0$). Generally, TNG predicts the highest {\HI} fractions in voids, while CHILES finds the lowest {\HI} fraction in voids.

\subsubsection{Specific Star-Formation Rate}\label{sec:ssfr_results}

\quad Using the procedures defined above, we summarize our result for relative sSFR in Figures \ref{fig:ssfr_z} and \ref{fig:ssfr_dfil}. We report our measurements as ``relative" sSFR because we take the difference of the measured sSFR and the average sSFR of each redshift range to account for the fact that the average sSFR of the universe changes significantly as a function of redshift \citep{madau14}. In Figure \ref{fig:ssfr_z}, we show the deviation from the average sSFR as a function of the redshift. The observations are represented by black markers, while the simulation is shown as a red-shaded region. We divided the data into three panels that correspond to three different cosmic web environments that we studied. In Figure \ref{fig:ssfr_dfil}, we present the deviation from the average sSFR as a function of D$_{fil}$ and categorize our measurements into two panels. The top panel displays the observations, while the bottom panel demonstrates the simulation results.

\quad Our observational results differ the simulation, but less than the deviations in the {\HI} fraction we noted in section~\ref{sec:fhi_results}. In the filament cores and boundaries, the observed deviation from mean sSFR is within $\sim$10\% of the predicted value at $z\leq0.5$, including very good agreement at $z\sim0.1-0.3$ in filament boundaries (Figure~\ref{fig:ssfr_z}). In voids, however, the relative star formation in CHILES is significantly lower than predicted by TNG. The relative sSFR declines from lower to higher $z$ in filament cores in both simulations and observations. In the filament outskirts, simulations predict a small rise in relative sSFR with increasing $z$, while observations show a small decline. In the voids, the difference is more noticeable; relative sSFR rises substantially with redshift in TNG but clearly falls with redshift in CHILES. The dependence of filament distance in TNG is such that at $z=0.5$, there is progressively higher star formation activity further away from filaments and at $z=0$, there is perhaps a slight increase in star formation from filament cores to voids (Figure~\ref{fig:ssfr_dfil}). In contrast, CHILES shows a rise in relative sSFR from filament cores to outskirts, followed by a decline in voids, in both redshift bins. Although there is a possibility of a correlation between the {\HI} gas and the deviation from the mean sSFR, we do not observe a similar tracking between excess or dearth of {\HI} and deviation from the mean sSFR. In line with the results for {\HI} fraction, TNG predicts the most star formation in voids, while CHILES measures the least.

\quad We note here that the observed SFR we adopt are from 1.4 GHz continuum luminosity, which traces star formation over fairly long timescales of $\sim$100 Myr \citep[e.g.,][]{Hao11}. In contrast, we choose instantaneous SFRs from TNG which are not directly equivalent to the various indicators of star formation (such as nebular emission lines or continuum emission) that are scaled to SFR \citep[see, e.g.,][and references therein]{KE12}. To test the effect of using time-averaged instead of instantaneous SFRs, we use measurements from the TNG catalogs \citep[as presented in, e.g.,][]{Donnari19}. We find that the qualitative aspects of the cosmic web and redshift dependence of deviation from the average sSFR do not change with differences in the adopted SFR definition. In particular, regardless of instantaneous, 100 Myr-averaged, or 1000 Myr-averaged SFRs adopted, the galaxies in filament cores and voids have the least and most star formation, respectively, while those in filament outskirts have approximately average levels of star formation (i.e., no deviation from the mean), at redshifts $z\lesssim1$. Slight quantitative differences are seen at $z\gtrsim1.5$ between 1000 Myr-averaged SFR and 100 Myr-averaged SFR; for example, a drop in relative sSFR with $z$ in the voids for the latter and a rise with $z$ in the latter. Overall, the main difference between the different adopted SFRs is simply in the $\pm1\sigma$ range in the deviation from the mean sSFR -- averaging SFR over longer timescales reduces the dispersion on the mean values -- and our main conclusions on the comparison between TNG and CHILES hold.

\section{Discussion} \label{sec:disc}

\quad To better understand the measurements presented in Section~\ref{sec:results}, it is essential to place them in the context of previous observations and theoretical studies. However, it is a challenging task, as many previous observational and theoretical works have used different experimental designs and approaches to quantify the impact of cosmic environments on galaxy properties. One crucial factor to consider is how large-scale cosmic environments are defined. Although there are many similarities in the spatial distribution and statistics of filaments, \citet{libeskind18} reported some large differences as well, including in the mass/volume fractions and halo mass functions. Moreover, different studies are based on galaxies selected using different criteria (stellar mass bins, local densities, color cutoffs, etc.). We must keep these differences in mind when evaluating the qualitative similarities and distinctions between our findings and those of other observational and theoretical works.

\subsection{Disagreements between simulations and observations}\label{sec:disagree}

\quad We uncovered significant discrepancies between the observational results and the theoretical predictions presented in this paper. The first major one is in the dependence of {\HI} and star formation on the cosmic web environment. TNG predicts rising or roughly constant average {\HI} and star formation from denser to diffuse environments, but CHILES shows the lowest {\HI} and star formation in the least dense regions (voids). Another source of strong tension is in the evolution of {\HI} gas fraction and star formation from $z\!\sim\!0.1$ to $z\!\sim\!0.5$ in the same regions of the cosmic web. The CHILES data show no evolution in the mean {\HI} fraction in the filament interiors and a very small decrease in redshift for the filament outskirts, in contrast to TNG, which predicts a strong increase in all cosmic web environments. Across virtually every bin in redshift and {\dfil} in which we compare observations and simulations, TNG overpredicts the observed cold gas as seen in Figures \ref{fig:fHI_z} \& \ref{fig:fHI_dfil}.   

\quad Simulations such as TNG are tuned to reproduce selected observations such as the stellar mass function. If the feedback from stars and AGN is increased, the gas in galaxies would be expelled to much larger radii and/or heated much more, which would reduce the amount of cold {\HI} gas to be more in agreement with observations. But this would likely give rise to strong tension with other properties such as the stellar mass function, {\HI} mass function, or column density distribution of {\HI} absorbing systems. We also estimated the {\HI} fraction within a much larger radial extent of several hundred kpc from the galactic centers and found that the dependence of {\HI} fraction as a function of filament-centric distance or the redshift does not change. The same is true when we consider {\it total} bound gas fraction as opposed to just {\HI} fraction \cite[see also, e.g.,][]{Hasan23}.
We discuss in detail the impact of different physical prescriptions, numerical methods, resolutions, etc., on the predictions of various simulations in section~\ref{discusstheory}.

\quad There are several uncertainties associated with the measurements of {\HI} gas fraction in TNG presented in this paper. One major uncertainty is in the details of the UV flux incident upon the gas. The photodissociation of {\Hmol} molecules into atomic {\HI} via Lyman-Werner (LW) radiation affects the molecular fraction of neutral hydrogen \citep[e.g.,][]{Elmegreen93,Ricotti01,Yoshida03}, which in turn is sensitive to the assumed form of the UVB radiation. Estimating the LW flux includes an assumption of escape fraction of radiation from molecular gas clouds, optically thin propagation through the ISM, with an (often simplified) assumed geometry, all of which carry uncertainties. \citet{Diemer18} showed that extremely low or high escape fractions can change the {\Hmol} fraction by up to a factor of 3, which could account for the discrepancy in the {\HI} fraction between simulation and observation, but the poorly-understood physical mechanisms involved in the estimation of escape fraction prevent a stronger conclusion from being drawn. \citet{Diemer18} also found that the UV flux is well-matched between different adopted UVB models at $z\lesssim1$, and that the LW flux is very low at $z=0$. Crucially, \citet{Gebek23} showed that modeling realistic UV radiation fields including dust attenuation in the ISM can significantly affect the {\HI} distribution of individual galaxies, but not that of ensemble statistics such as {\HI} mass functions.

\quad Another source of uncertainty is in the neutral gas fraction itself (i.e., {\HI}+{\Hmol}), which is affected by ionizing radiation from young stars or AGN. \citet{Rahmati13}, for example, reported a significant drop in neutral gas fraction with increasing irradiation from young stars, which would affect the dense star-forming gas cells in TNG. However, this effect is more pronounced for high-redshift galaxies, and here, too, the escape fraction of photons is highly uncertain \citep[e.g.,][]{Wise09,Finkelstein19}. Generally, a higher escape fraction of LW radiation would result in a higher {\HI} fraction (due to a lower {\Hmol} fraction) and a higher escape fraction of ionizing radiation would result in a lower {\HI} fraction (due to a lower neutral fraction).

\quad On the observational side, one source of uncertainty is the existence of low-mass satellite galaxies contaminating the stacked CHILES {\HI} emission measurements as a source of confusion. \citet{Jones16} generated an analytical model to describe the contribution of {\HI} mass from confused sources to stacked spectra in {\HI} surveys. They showed that at the native CHILES resolution of 5$\arcsec$, confusion noise in the stacked measurements would be on the level of $\approx10^{8}~{\Msun}$ at $z=0.45$, which is an order of magnitude below our presented measurements. Even if confusing sources were dominant in the stacked spectra, accounting for this would actually reduce the {\HI} fraction and star formation, which would increase tension with simulations, instead of alleviate them.

\quad Cosmic variance and small number statistics might be causes of concern for our observations. The 40$\arcmin$ length image of CHILES corresponds to $\lesssim$15 Mpc in physical size at $z=0.1-0.5$ and $\sim300-450$ Mpc in the redshift direction. This volume is about one order of magnitude smaller than the $\sim$110.7 Mpc length of a TNG100 simulation box. Cosmic variance might be affecting the statistics of higher mass galaxies in our sample, but we show in the following section how the statistics of high-mass galaxies are in better agreement between theory and observations than those of low-mass galaxies. CHILES may not be adequately sampling low-density regions, which could explain the non-detection of a significant {\HI} emission in the voids at $z\sim0.3-0.5$ (Figure~\ref{fig:StackSpectra}) and the strong discrepancies between predicted and observed {\HI} fractions and relative sSFR. Additionally, the CHILES statistics are computed from $\sim$70-300 galaxies per stack in redshift and {\dfil} bins, whereas the TNG statistics are based upon a thousand or more galaxies in each bin, thus larger by over one order of magnitude.

\begin{figure*}[h!t!]
\begin{center}
\includegraphics[width=\textwidth]{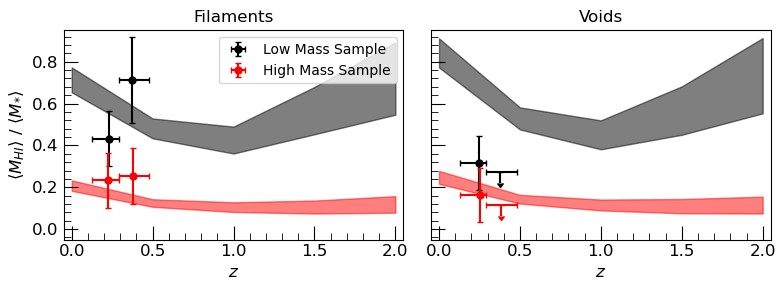}
\caption{The mean {\HI} gas fraction as a function of redshift in two different cosmic web environments: filaments (left; $0-3$ Mpc) and voids (right; $3-20$ Mpc). The points with error bars correspond to the observed measurements, and the shaded regions correspond to the results from TNG100. The low-mass and high-mass samples are comprised of galaxies in the stellar mass range 10$^{9-10}$ M$_{\odot}$ and 10$^{10-12.5}$ M$_{\odot}$, respectively. There is much better agreement for high-mass galaxies than for low-mass galaxies, and in filaments than in voids.}
\label{fig:fHI_MassBins}
\end{center}
\end{figure*}

\subsection{Comparison to Other Observations}\label{sec:obsdisc}

\quad There have been multiple studies investigating the role of the cosmic web in the cold gas content of galaxies for the local universe \citep{kleiner17,odekon18,Hoosain24}, that use data from wide-field local {\HI} surveys providing them with powerful statistical samples. As well as some recent work expanding this analysis to intermediate redshifts \citep{sinigaglia24}. Each of these studies has slightly different methodologies, samples, results, and interpretations, and in the following section, we will examine how our observed results fit into the larger context of the literature.

\quad We start by briefly describing the work done for the local universe. \citet{kleiner17} used the LSS reconstructed from the 6 degree Field Galaxy Survey \citep{jones09} and the {\HI} data from the {\HI} Parkes All Sky Survey \citep{barnes01} to investigate differences in the {\HI} gas and stellar mass fractions in different LSS environments. Via the {\HI} gas fraction as a function of the fifth nearest neighbor density for a sample of galaxies located close to filaments (D$_{fil}$ $<$ 0.7 Mpc) and away from filaments (D$_{fil}$ $> $5 Mpc), the authors reported that the gas fraction differs only for galaxies with a stellar mass greater than 10$^{11}$M$_{\odot}$, and galaxies near filaments have a higher gas fraction. Meanwhile, \citet{odekon18} defined the structure in the Sloan Digital Sky Survey \citep{york00} field using an adaptation of the technique by \citet{alpaslan14} and {\HI} data from 70\% of the ALFALFA survey \citep{giovanelli05}. They investigated {\HI} deficiency, which is the deviation of the measured {\HI} mass from a predicted {\HI} mass \citep{giovanelli83} as a function of the distance from the nearest filament. They found that galaxies are proportionately less {\HI} deficient with increasing distance from filaments, but that this trend flattens out for galaxies that are more than $\sim$3~Mpc away from filaments. Similar to their results were those reported recently by \citet{Hoosain24}, who investigated the gas fraction ($1.4{\MHI}/{\ms}$) of nearby galaxies in the RESOLVE survey and the ECO catalog \citep{Eckert16,Stark16} as a function of the cosmic web reconstructed by {\disperse}. They found that galaxies of masses $8.5<{\logms}<10.5$ are more gas-poor close to filaments, with the trend being strongest for low- and intermediate-mass galaxies in groups and isolated low-mass centrals. Recently, using data from the MIGHTEE-HI project \citep{maddox21} at redshift $z\approx0.37$, \citet{sinigaglia24} stacked {\HI} emission in galaxies with a focus on their correlation with the cosmic web. Specifically, \citet{sinigaglia24} use photometric redshifts from COSMOS2015 \citep{laigle16} as inputs to a Hessian Matrix based classification method that allow them to distinguish different cosmic web environments and geometrical features of the cosmic web (we refer the reader to Sections 3.1 $\&$ 3.2 in \citet{sinigaglia24} for a thorough description). The authors find that at $z\approx0.37$, (1) filament galaxies are the most {\HI} rich, (2) field galaxies (most similar to our ``void" galaxies) have slightly elevated {\HI} levels, and (3) galaxies residing in knots are {\HI} deficient. These results in conjunction with their probes of the local environment, lead them to conclude that {\HI} content depends on the large-scale environment of a galaxy, such that intermediate-density environments, filaments and outskirts of DM halos, host the most {\HI}-rich galaxies.

\quad On the surface, the results of \citet{kleiner17}, \citet{odekon18}, and \citet{Hoosain24} appear to contradict one another. While \citet{kleiner17} found {\HI} enrichment closer to the filaments, both \citet{odekon18} and \citet{Hoosain24} found {\HI} deficiency close to filaments. However, \citet{kleiner17} binarily focused on filament vs. non-filament, while \citet{odekon18} studied properties as a function of distance to filaments and smaller filaments, referred to as tendrils, within the underdense regions. \citet{odekon18} also pointed out that the statistically significant difference in the \citet{kleiner17} study was found for galaxies with a stellar mass greater than 10$^{11}$M$_{\odot}$, a range for which \citet{odekon18} had very few galaxies, as was also observed in \citet{Hoosain24}. Moreover, differences between results could also result, in part, due to the use of different structure-defining algorithms. Similar to our work, \citet{kleiner17} and \citet{Hoosain24} reconstructed the cosmic web using {\disperse}, while \citet{odekon18} did so using the MST algorithm \citep{alpaslan14}.

\quad To begin to place our work within this context, we will compare our work to that presented in \citet{kleiner17}. In order to draw the best comparison possible given observational constraints, we compare the {\HI} gas fractions for galaxies in two different stellar mass bins, a low mass bin (10$^{9-10}$ M$_{\odot}$) and a high mass bin (10$^{10-12.5}$ M$_{\odot}$), in two different environments, associated with filament ($0-3$ Mpc) and void ($3-20$ Mpc) galaxies, for the same two redshift bins that we have been studying throughout this work. In Figure \ref{fig:fHI_MassBins}, we illustrate this analysis and find, first, low-mass galaxies are more {\HI} rich than high-mass galaxies in both CHILES and TNG; second, galaxies associated with filaments are more {\HI} rich than those associated with voids in CHILES. Galaxies closer to filaments being more {\HI}-rich than further away is similar to the results of \citet{kleiner17}, but we find this to be true for both mass bins in CHILES, whereas \citet{kleiner17} finds this in only the most massive stellar mass bin. Interestingly, Figure \ref{fig:fHI_MassBins} shows that the simulations appear to do a better job at predicting the {\HI} content of galaxies of higher stellar mass than lower stellar mass, including quantitatively matching the {\HI} fraction in filaments and capturing the redshift-evolution in both filaments and voids. The simulations agree much better with the data in the filaments, but continue to overpredict the {\HI} in voids. When comparing our results to \citet{odekon18} and \citet{Hoosain24}, we seem to have conflicting results insofar as our sample of galaxies become more {\HI} rich, in both redshift bins, as the galaxies live closer to the filaments. 

\quad While we conclude that our results do not always agree with the aforementioned studies, we warn the reader against over-interpretation as different experimental designs, galaxy samples, and cosmic web-defining algorithms were implemented in each of the different analyses. Lastly, we compare our results to the data points presented in \citet{sinigaglia24} for redshifts, $z\approx0.37$. The most straightforward points to compare are the ones they label field and filaments show that at these redshifts, galaxies closer to filaments are more {\HI} rich than the field (void) galaxies. This is also the result that we recover for galaxies in a similar redshift range ($0.3<z<0.48$) for both the total ensemble of galaxies (see Figure \ref{fig:fHI_z}) and galaxies segregated by mass (see Figure \ref{fig:fHI_MassBins}). However, we remind the reader that there is substantial overlap between the MIGHTEE and CHILES field and that our agreement with \citet{sinigaglia24} offers confirmation that our differing methods reproduce similar results, as opposed to an independent validation.

\quad The fact that we do not find significant evolution in this redshift range, but do observe a difference for our two stellar mass bins is in good-agreement with stacking experiments in the literature. Specifically, stacking results from \citet{bera19} show little-to-no evolution of the HI-stellar mass scaling relationship out to redshifts at approximately 0.35. Additionally, in \citet{LuberStacking}, we find that for the total ensemble of blue galaxies there is no evolution in total gas content. However, in \citet{LuberStacking}, we do find differences in the evolution of HI content for blue galaxies when we segregate by mass, and indeed in this work we find differing HI contents for our two different mass bins. Given that \citet{chowdhury22a} finds significant evolution of HI content at redshifts $\sim$1, it must be that this evolution, and its probable dependence on environment, occurs in earnest beyond the redshifts we probe in this work.

\subsection{The Theoretical Context} \label{discusstheory}

\subsubsection{The Gas Fraction in Simulations}

\quad Reproducing the observed cold gas content of galaxies is challenging in hydrodynamical cosmological simulations. The primary reason is that simulations such as EAGLE, TNG, and SIMBA are calibrated to match observed stellar and star formation properties of galaxies but not (cold) gas properties. The differences in numerical techniques and physical prescriptions of galactic feedback physics result in significant disagreements in the properties of cold gas between these simulations \citep{Dave20}, and subsequently, this can lead to varying degrees of agreement of a given simulation with observations.

\quad Different models employ varying subgrid physics implementations that determine star formation feedback (crucial for lower masses) and AGN feedback (important for higher masses). For example, in EAGLE, SF-driven outflows are implemented by injecting thermal energy into gas, which heat up the ISM and CGM. \citet{Dave20} showed that this generates a lower {\HI} gas fraction in low-mass galaxies in EAGLE, compared to simulations such as TNG and SIMBA in which kinetic energy is deposited into outflows that are decoupled from the ambient ISM. In TNG, the AGN feedback is in the form of thermal energy injection at high BH accretion rates, which can heat the gas, and kinetic energy injection at low BH accretion rates, which can expel the gas far from galaxies \citep{Zinger20}. When the central BH mass of a galaxy is ${\log (M_{\mathrm{BH}}/{\Msun})\geq8.2}$, kinetic AGN feedback dominates the thermal feedback, rapidly clearing out the gas supply and leading to quenching of galaxies \citep{Terrazas20}. In contrast, EAGLE has only a thermal feedback mode where the thermal energy imparted by the BH is proportional to the BH mass. Additionally, EAGLE uses the smoothed particle hydrodynamics code {\sc GADGET} \citep{springel05}, while TNG uses the moving-mesh refinement hydrodynamics code {\sc AREPO} \citep{Springel10}. Despite differences in subgrid physics and numerical schemes, both simulations qualitatively produce the same trend of rapid quenching by gas expulsion for high-mass galaxies \citep{Davies20}.

\quad According to \citet{Dave20}, TNG overpredicts the $z=0$ {\HI} gas fraction in ${\logms}\gtrsim10.5$ galaxies, but reproduces the {\HI} fraction at lower masses. The study by \citet{Diemer19}, which used the same catalogs as the current paper, reported that TNG agrees with the observed median {\HI} at all stellar masses in the local universe. This apparent disagreement between the two studies is caused by the much smaller aperture used by \citet{Diemer19} to measure {\HI}, which misses the lower {\HI} gas content further from galaxies. However, TNG100 overestimates, by a factor of $\sim$2, the overall cosmic abundance of {\HI} observed by 21 cm surveys at $z\approx0$, which could explain some of the overprediction of {\HI} relative to CHILES as we find here. In contrast, the fiducial EAGLE model generates a lower-than-observed {\HI} content at $z=0$, possibly due to overheating of gas in galaxies. This tension, \citet{Dave20} found, can be alleviated with a higher-resolution run of the simulation. They concluded that models with kinetic winds decoupled from thermal feedback (such as that in TNG) are less sensitive to resolution effects.
Similarly, \citet{Diemer19} showed that TNG50 produces {\HI} fractions in agreement with TNG100, at least for the mass ranges we examine in this paper.

\subsubsection{The Relationship of Gas Content With Environment}

\quad There have only been a few studies that have measured the correlation between the gas fraction of galaxies and the LSS environment in hydrodynamical cosmological simulations, yielding mixed results. \citet{Singh20} reconstructed the cosmic web from $\log(\ms/{\Msun})>9$ galaxies in EAGLE at $z\simeq0.1$ using {\disperse}. They found that the gas fraction ${\mgas}/{\ms}$ and SFR increase with {\dfil} at ${\dfil}\gtrsim0.5$~Mpc and decrease with {\dfil} at ${\dfil}\lesssim0.5$, interpreting this as possible evidence of intra-filamentary gas condensing onto galaxies at the cores of filaments. Using different methods, \citet{RG22} found that the {\HI} fraction in EAGLE is lower in $\log(\ms/{\Msun})\sim9.5-10.5$ galaxies residing in voids than those in denser environments, while the values are comparable for more massive galaxies in different environments. Using the {\disperse} framework, \citet{Bulichi24} found a reduction in {\HI} gas fraction and star formation in galaxies with increasing proximity to filaments in SIMBA at $z\lesssim1$. Finally, \citet{Zakharova24} applied {\disperse} to both observational data and semi-analytical models to find that galaxies in filaments possess less {\HI} than those in the field.

\quad For TNG, \citet{Hasan23} utilized {\disperse} to show that the total gas fraction (defined as ${{\mgas}/({\mgas}+{\ms})}$) increases with increasing {\dfil} at low redshift. They found no evidence of an increase in gas fraction near the filament spines, and a smooth decline in gas fraction from $z=2$ to 0 for a given cosmic web environment, while our results here suggest a rise in {\HI} fraction at $z<1$. This difference could be due to a combination of factors. Due to the dominant AGN contribution of the UVB at $z\leq3$, the {\HI} photoionization rate drops significantly from $z\sim1.5-2$ to $z=0$ \citep[e.g.,][]{BB13,Khaire19,Kulkarni19,FG20}. Less  ionization of neutral hydrogen could explain the increase in average {\HI} fraction, but not gas fraction, from $z=1$ to 0. The {\Hmol} gas density is also known to decrease in the $z=1\rightarrow0$ interval, but simulations such as TNG tend to underpredict the observed {\Hmol} density  \citep[e.g.,][]{Popping19,PH20}. Additionally, \citet{Hasan23} do not impose an upper limit on the distance from a galactic center as we do here, instead considering all the gas bound to the CGM and ISM. Our restriction to blue galaxies also differentiates our sample from those used in the aforementioned studies.

\quad Several other factors than simply a galaxy's location relative to the LSS may play an important role in regulating its gas content. For instance, \citet{zhu22}, who analyzed a hydrodynamical simulation run with the RAMSES adaptive mesh refinement code \citep{teyssier02}, reported that cold ($T<10^{5.5}$~K) gas accretion onto DM halos is dependent on the widths of filaments supplying the gas, which they measured using a Hessian matrices (different from {\disperse}). Filaments that were $>3~h^{-1}$~Mpc in width had $\approx2-3$ times lower cold gas fraction than those with width $<3~h^{-1}$~Mpc. At $z<1$, the cold gas accretion rate onto halos with $\log(\mh/{\Msun})\lesssim12$ was lower in thicker filaments. Recently, \citet{Lu24} quantified filament boundaries by the existence of virial shocks and virial equilibrium per unit length in the DM halo potential well, yielding consistent results. They predicted that thicker filaments have higher virial temperatures, which would imply lower cold gas fractions. Some authors such as \citet{GE22} have also shown that the length of filaments is correlated with their gaseous properties, which can impact the amount of gas that is accreted onto galaxies. Additionally, \citet{Song21} reconstructed the cosmic web at $z\sim2$ in the Horizon-AGN simulations \citep{Dubois14} using {\disperse} and found that the outer region of a halo is more gas-rich at lower {\dfil}, but that the galaxy itself is less gas-rich. They suggested that this inefficiency of gas transfer from outer to inner regions of halos is due to the coherent, high angular momentum supply of gas from filaments to outer halos, which hinders gas transfer down to the galaxy centers.

\quad Unfortunately, the {\disperse} outputs are point sets, making it difficult to measure the width of the filaments using this method. In a separate work, a new method based on the biological {\it slime mold} organism \citep{Burchett20,Elek22} was leveraged by \citet{Hasan24} to reconstruct the cosmic web and understand the effect of filamentary density on galaxy evolution. A crucial finding of this work is that there is a strong correlation between the gas fraction and star formation of galaxies and the density along the filament segment closest to a galaxy at $z\leq1$. This suggests that the diversity in the properties of filaments, beyond simply the Euclidean distance to filament spines, is very important to consider when attempting to understand how filaments affect galaxy formation phenomena. The filament-to-filament diversity underlies the different physical mechanisms that can act on the gas reservoirs of galaxies. In other words, we do not expect filaments to have a monolithic effect on galactic gas supply and star formation.

\quad Several theoretical frameworks have been proposed to explain how the gas supply of galaxies is tied to the cosmic web. One such framework is the so-called "cosmic web detachment" (CWD) model \citep{cwd}, in which galaxies are cut off from their primordial gas supply through events such as major mergers, satellite galaxy accretion, and interaction with cosmic web filaments. These processes result in gas starvation that can quench galaxies independent of internal mechanisms. Starvation is a slow phenomenon that quenches galaxies on longer timescales \citep[few Gyr on average; e.g.,][]{Peng15} than rapid gas removal phenomena such as ram pressure stripping \citep[e.g.,][]{gunn72} or tidal stripping \citep[e.g.,][]{Marasco16}, which are applicable in massive group/cluster-sized halos (our galaxy samples have virtually none of these massive halos). Furthermore, \citet{BL13} proposed that large fractions of present-day low-mass galaxies may lose a significant fraction of their gas via ``cosmic web stripping'', i.e., through interaction of galaxies with the cosmic web itself \citep[see also, e.g.,][]{Herzog23}. At $z\!\sim\!2-5$, \citet{Pasha23} found that the outskirts of sheets can act similarly to hot halo environments in suppressing gas accretion and quenching dwarf galaxies.

\quad TNG predicts that the {\HI} gas fraction drops in galaxies residing in higher-density environments (${\dfil}\leq2$~Mpc) at $z=0-0.5$ (Figure~\ref{fig:fHI_dfil}). This could be due to CWD events that restrict the cold gas supply to galaxies as they move closer to filaments. This detachment from the cold supply happens when galaxies cross or accrete onto a cosmic filament that detaches them from the much thinner primordial filaments that can supply cold gas. According to the \citet{cwd} model, for halos with masses $\log({\mh}/{\Msun})=11.5-12.5$ (roughly equivalent to stellar masses $\log({\ms}/{\Msun})=9-10.5$ from stellar-to-halo mass relations; e.g., \citealt{Behroozi19}), CWD events peaked around $z\approx2$ and declined at later times, until the present day. This is in line with the decrease in {\HI} content observed in TNG, which decreases with the redshift from $z\sim2$ to $z\sim0.5$. However, the increase in {\HI} from $z\sim0.5$ to the present day in TNG contradicts the CWD model, while CHILES observations of roughly constant, or slightly declining, {\HI} fraction at lower redshifts are consistent with the model.

\quad An alternate explanation to CWD is filamentary condensation, which proposes that filaments of a given line mass are prone to cooling, which can feed halos that reside near the spine of the filament \citep{Birnboim16}. Their models predict that condensation from unstable filaments can feed halos with ${\mh}\sim10^{11.5}-10^{14.5}$ at $z\sim0-0.5$ (this range of masses decreases with increasing redshift in their model). This approximately encompasses galaxies in our stellar mass range of ${\logms}=9-10.5$, and could help explain the physics behind the high {\HI} fraction in CHILES galaxies at ${\dfil}<2$~Mpc. However, it has been shown that cold streams are less likely with decreasing redshifts to penetrate the hot CGM of massive halos at lower $z$ due to hydrodynamical instabilities and inefficient cooling in the turbulent mixing zones at the boundaries between the streams and the CGM \citep[e.g.,][]{Mandelker20a,Daddi22a,Daddi22b}. 

\section{Summary} \label{sec:conc}

\quad In this work, we measure the average {\HI} gas fraction and relative sSFR of blue galaxies (NUV-r=-1 - 3) as a function of their location in the cosmic web and redshift. We do this for both observations, radio interferometric data from CHILES and ancillary data available in the COSMOS field, and the TNG100 cosmological simulation. For both the observed and simulated galaxies, we used the scale-free topological structure finding algorithm DisPerSE to define the filaments that comprise the cosmic web. Using only blue galaxies and a stellar mass in the range ${\logms}=9-11.5$, we then compare the measurements made from the observations and the TNG100 simulation. Our main results are as follows.

\begin{itemize}
\item CHILES observations indicate that the mean {\HI} gas fraction of galaxies decreases as a function of distance from the spine of cosmic web filaments for both the high redshift ($0.3<z<0.48$) and low redshift ($0.13<z<0.29$) bins (Figure~\ref{fig:fHI_dfil}). There is almost no increase in the {\HI} fraction in filament cores and outskirts from $z\sim0.1$ to $z\sim0.5$ within the error bars, but there is a substantial decline in {\HI} in voids in this time (Figure~\ref{fig:fHI_z}).

\item Our measurements from the TNG100 simulation, meanwhile, suggest a increase in the {\HI} gas fraction from $z=0.5$ to 0, in all cosmic web environments (Figure~\ref{fig:fHI_z}). At $z=0.5$, the {\HI} fraction increases monotonically with filament-centric distance, while at $z=0$, {\HI} fraction is roughly the same in filament cores and outskirts, and only slightly larger in voids (Figure~\ref{fig:fHI_dfil}). In TNG100, the {\HI} fraction reaches a minimum at $z=1$ for all cosmic web environments, and then increases again at $z>1$. 

\item CHILES observed an increase in the deviation from the mean sSFR at a given redshift, i.e., relative sSFR, from filament cores to the outskirts, and then a decrease to a minimum in the voids -- both for the low and high redshift bins (Figure~\ref{fig:ssfr_dfil}). There is a small decrease in relative sSFR within the same cosmic web environment with increasing redshift (Figure~\ref{fig:ssfr_z}).

\item For TNG100, we found a complex evolution of relative sSFR with respect to redshift in each of our cosmic web bins, each showing different evolutionary tracks. At $z=0.5$, we observed an increase in differential sSFR in TNG with distance from filaments but little change at $z=0$ (Figures~\ref{fig:ssfr_z} and \ref{fig:ssfr_dfil}). 

\item We found that simulations and observations agree that low-mass (M$_{*}$ = 10$^{9-10}$ M$_{\odot}$) blue galaxies are more {\HI}-rich than high-mass (M$_{*}$ = 10$^{10-12.5}$) galaxies in the same environment and redshift. There is reasonably good agreement when considering filamentary regions ($0-3$ Mpc) rather than voids ($3-20$ Mpc), where TNG still predicts too much {\HI} relative to CHILES. Interestingly, TNG reproduces the {\HI} fraction and its redshift-evolution for high-mass blue galaxies quite well, but disagrees with low-mass blue galaxies considerably.

\item We discussed these results in the context of the successes and failures of cosmological simulations and their treatment of baryonic physics. Additionally, we examined our findings in the context of the CWD model of galaxy evolution, and find our results align well with this theoretical framework. 

\end{itemize}

\quad Comparing observations and simulations of galaxies across a wide range of parameters is crucial to test theories of galaxy evolution. In this work, we compared the {\HI} gas fraction for galaxies in different LSS environments over a significant range of redshifts. We found significant discrepancies between simulations and observations regarding the reported measurements of {\HI} gas fraction and differential sSFR. However, although the CHILES data provide an initial insight into the redshift evolution of the {\HI} gas fraction, our conclusions are limited due to the sensitivity of the current measurements. More sensitive measurements are required to confirm our results and interpretation. Future astronomical surveys will make experiments like this easier and more informative. With higher galaxy counts, larger angular areas, deeper redshift coverage, and higher sensitivity, these surveys will be able to produce more stacked {\HI} measurements and compare individually detected galaxies across an increased redshift range. Forthcoming advances in observations should clarify and resolve discrepancies among different observational measurements in the literature, enable rigorous tests of galaxy formation models, and advance our understanding of how large-scale cosmic-web environments affect the expected gas content of galaxies. 

\begin{acknowledgments}

\vspace{0.5cm}

\quad We greatly appreciate and acknowledge the thoughtful comments from the anonymous referee that helped improve the presentation of results, scientific discussion, and clearness. We acknowledge insightful discussions with R. Dav{\'e}, J. Tumlinson, S. Tonnesen, J. Woo, K. Finlator, N. Kanekar., and S. Mishra. FH acknowledges Z. Edwards and D. Hellinger for their support in setting up and running {\disperse}. The National Radio Astronomy Observatory is a facility of the National Science Foundation operated under cooperative agreement by Associated Universities, Inc. This research used resources from the National Energy Research Scientific Computing Center, a DOE Office of Science User Facility supported by the Office of Science of the US Department of Energy under Contract No. DE-AC02-05CH11231 using the NERSC award HEP-ERCAP0024028. The IllustrisTNG simulations were undertaken with compute time awarded by the Gauss Centre for Supercomputing (GCS) under GCS Large-Scale Projects GCS-ILLU and GCS-DWAR on the GCS share of the supercomputer Hazel Hen at the High Performance Computing Center Stuttgart (HLRS), as well as on the machines of the Max Planck Computing and Data Facility (MPCDF) in Garching, Germany. The CHILES survey was partially supported by a collaborative research grant from the National Science Foundation under grant Nos. AST—1412843, 1412578, 1413102, 1413099, and 1412503. FH and JNB are supported by the National Science Foundation LEAPS-MPS award $\#2137452$. DN is supported by the NSF AST-2307280 grant. NM acknowledges support from the Israel Science Foundation (ISF) grant 3061/21 and US-Israel Binational Science Foundation (BSF) grant 2020302. DJP and SK gratefully acknowledges support by the South African Research Chairs Initiative of the Department of Science and Technology and the National Research Foundation. Julia Blue Bird was a Jansky Fellow of the National Radio Astronomy Observatory.

\end{acknowledgments}

\facilities{EVLA \citep{evla}, WVU High Performance Computing}

\software{{\sc astropy} \citep{astropy}, {\sc casa} \citep{casa}, {\disperse} \citep{sousbie11a}, {\sc scipy} \citep{scipy20}}

\bibliography{references}{}
\bibliographystyle{aasjournal}

\end{document}